\documentstyle[aps,epsfig,float,twocolumn,prl]{revtex}
\newcommand{\be}{\begin{equation}}
\newcommand{\ee}{\end{equation}}
\begin{document}
\twocolumn[\hsize\textwidth\columnwidth\hsize\csname @twocolumnfalse\endcsname
\draft
\title{Mean-Field and Anomalous Behavior on a Small-World Network}
\author{M. B. Hastings}
\address{
Center for Nonlinear Studies and Theoretical Division, Los
Alamos National Laboratory, Los Alamos, NM 87545,
hastings@cnls.lanl.gov }
\date{April 21, 2003}
\maketitle
\begin{abstract}
We use scaling results to identify the crossover to mean-field behavior
of equilibrium statistical mechanics models on a variant
of the small world network.  The results are
generalizable to a wide-range of equilibrium systems.  Anomalous
scaling is found in the width of the mean-field region, as well
as in the mean-field amplitudes.  Finally,
we consider non-equilibrium processes.
\vskip2mm
\end{abstract}
]
The appropriate description for many complex real-world systems
is as a network\cite{nr}, a general connection of nodes and vertices
which need not have the structure of a regular lattice.  The small-world
network model combines both long-range and short-range aspects,
and interpolates between regular lattices and random graphs.
This model\cite{sw}, in which a regular lattice is modified by
either randomly ``re-wiring" links or else by simply randomly
adding long-range links,
has become a standard model of real-world networks.  It incorporates
some notion of locality, as most of the links remain the same as that of
the original underlying lattice.  Yet it also includes the 
``small world effect", that the average path length between sites on the
network scales only as the logarithm of the network size.

Cooperative behavior on networks is a major topic of interest.
Studies of equilibrium statistical on these systems have shown rich
behavior\cite{eq,eq2,eq3,eq4}.  
Work\cite{gdm,l} on networks with long tails in the probability 
distribution of the coordination number has found alternatives
to mean-field behavior.   We will instead study the behavior on small-world
systems.  The technique will be generically applicable to
a wide-range of problems.
We show the existence of mean-field behavior, albeit with
anomalous exponents describing the width of the critical region and
various mean-field amplitudes.  These anomalous exponents can
complicate the interpretation of numerical data.  Finally, we will
consider a non-equilibrium case, describing the relaxation to a stationary
state via a branching process.

The presence of quenched randomness makes the small-world model difficult
to treat analytically.  We instead consider a different, but related model,
which lacks quenched randomness, considering a variety of
equilibrium and non-equilibrium models.  We again start with a regular lattice
of $V$ sites in $d$ dimensions.  Rather than adding long-range links
with probability $p$, we give each site of the lattice a weak coupling,
of order $p/V$, to every other site in the lattice\cite{kardar}.
We will
refer to this as the long-range model.
We will find that many results can be obtained on this system by
combining mean-field with standard renormalization group techniques.
Later, we will compare this model to the standard small world model; we
will argue that in many cases the presence of quenched randomness in
the small world model is irrelevant, and thus scaling results for the
small world model can be obtained
from the long-range model.  It is interesting to compare to a related mean-field
solution of path lengths on a small work network\cite{mf}.  Finally,
we will show the long-range model provides an upper bound for the
free energy of the small world network model.

In some cases, the long-range model may also be more appropriate than the
usual small world model.  In the spread
of a disease, for example, people tend to spread the disease to
those geographically nearby (the regular lattice).  There is a chance
of a long-range spread of the disease.  However, this is not necessarily
due to fixed long-range links.  Rather, it is due to the random probability
that a given person travels a long distance, typically by air.  Thus,
a slight probability of long-range contact between any two people may
be a better description than a set of fixed long-range links.

{\it Equilibrium Statistics---}
We consider any equilibrium statistical mechanics model, such as
an Ising model, XY model, etc...  These models can be represented
by introducing a field $\phi(x)$, where $x$ labels lattice sites and
where $\phi$ has 1,2,... components, with
a partition function
\be
Z=\sum\limits_{\{\phi\}} \exp[-S[\phi]],
\ee
where $S=E/kT$ is a statistical weight for a configuration of
energy $E$ at temperature $T$.

For a model
on a regular $d$-dimensional lattice, $S[\phi]=S_{\rm local}[\phi]$,
a local function of $\phi$.  We refer to this as the local system.
We choose instead for the long-range model a statistical
weight with additional long-range couplings of strength $p/V$:
\be
\label{sl}
S[\phi]=S_{\rm local}[\phi]-\frac{p}{2V}\sum\limits_{x_1,x_2} \phi(x_1) \cdot
\phi(x_2).
\ee

Then, decouple the long-range interaction to find
\be
\label{zl}
Z=\int\limits_{-\infty}^{\infty} {\rm d}h\,\exp[-\frac{Vh^2}{2p}]
Z(h),
\ee
where
\be
Z(h)=
\sum\limits_{\{\phi\}} \exp[-S_{\rm local}[\phi]+h\sum\limits_x \phi{x}].
\ee
Here, $Z(h)$ is equal to the partition function of the local system 
in the presence of a magnetic field, $h$.

For $p$ large, the long-range interaction outweighs the short-range
interaction, and the system can be approximately
solved by mean-field theory.  We will instead consider precisely
the opposite case: when $p$ is small.
Then, the critical point of the long-range system, $\tilde T_c$,
is close to the critical point of
the local system, $T_c$.  Thus, we can use scaling laws for the
local system:
the magnetization, $m$, obeys $m(T=T_c,h)=A_m |h|^{1/\delta}$, defining
the critical exponent $\delta$.  
For $T>T_c$, the susceptibility
$\chi$ obeys $\chi(T,h=0)=A_{\chi}^+ |T-T_c|^{-\gamma}$.  In general,
we can write a scaling function: 
$m=h^{1/\delta} f((T-T_c)h^{-1/(\delta\beta)})$.  For 
$h>>(T-T_c)^{\delta\beta}$, we use the first result
$m(T=T_c,h)=A_m |h|^{1/\delta}$, while for
$h<<(T-T_c)^{\delta\beta}$, we use the second
$m(T=T_c,h)=A_{\chi}^+ h |T-T_c|^{-\gamma}-B_{\chi} h^3 
|T-T_c|^{-\gamma-2\delta\beta}$, where we have added the $h^3$ term in
the expansion of $m$.

The magnetization
is defined by $\partial {\rm ln} Z(h)/\partial h=mV$.  Thus,
\be
\label{zh}
Z(h)=Z(0) \exp[\int\limits_{0}^{h} {\rm d}h'\, m(h') V].
\ee

We start by considering the case of $T$ near $T_c$ so that
$h>>(T-T_c)^{\delta\beta}$.  Then, Eqs.~(\ref{zl},\ref{zh}) give
\be
Z=Z(0)\int\limits_{-\infty}^{\infty} {\rm d}h\,
\exp[-\frac{Vh^2}{2p}+A_mV h^{1+1/\delta}/(1+1/\delta)].
\ee
Since $V$ is taken large, we can use a saddle point to arrive at
$h=(pA_m)^{\delta/(\delta-1)}$, or 
\be
\label{mtc}
m=A_m(pA_m)^{1/(\delta-1)}.
\ee
The correlation length $\xi$ of the local system in the presence of this
field is proportional to $m^{-\nu/\beta}$, and hence diverges as
$p\rightarrow 0$.
The meaning of the correlation length is {\it not} that spins beyond this
length are uncorrelated.  Rather, it is that beyond this length the correlations
in the long-range system are controlled by the average field $m$, while below
this length, the fluctuations are important.  Thus, the correlation
function $\langle \phi(0)\phi(x) \rangle$ decays as a power law up to
the correlation length, and then asymptotes to a constant.

We have seen that at $T=T_c$, the system has a net magnetization,
and thus $T_c<\tilde T_c$.  To study the transition itself, we now consider
the second case, 
$h<<(T-T_c)^{\delta\beta}$.  Now, Eqs.~(\ref{zl},\ref{zh}) give
\begin{eqnarray}
\label{mfa}
Z=Z(0)\int\limits_{-\infty}^{\infty} {\rm d}h\,
\exp[-\frac{Vh^2}{2p}+A_{\chi}^+ V h^{2}|T-T_c|^{-\gamma}/2 \\ \nonumber
-B_{\chi} V h^4 |T-T_c|^{-\gamma-2\delta\beta}/4
].
\end{eqnarray}
Again using a saddle point, we find a critical point 
at $\tilde T_c$ given by
\be
\label{tct}
\tilde T_c-T_c=(pA_{\chi}^+)^{1/\gamma},
\ee
with $m=0$ for $T>\tilde T_c$.
Slightly below the critical point we find
$h=\sqrt{\tilde T_c-T}\sqrt{\frac{\gamma A_{\chi}^+}{B_{\chi}}}
(\tilde T_c-T_c)^{\delta\beta-1/2}$, and a magnetization given by
\begin{eqnarray}
\label{mtct}
m=\sqrt{\tilde T_c-T}A_{\chi}^+\sqrt{\frac{\gamma A_{\chi}^+}{B_{\chi}}}
(\tilde T_c-T_c)^{\delta\beta-1/2-\gamma} \\ \nonumber
=\sqrt{\tilde T_c-T}A_{\chi}^+\sqrt{\frac{\gamma A_{\chi}^+}{B_{\chi}}}
(\tilde T_c-T_c)^{\beta-1/2}
.
\end{eqnarray}
Thus, the magnetization behaves as
$m = \tilde A \sqrt{\tilde T_c-T}$, with
$\tilde A \propto p^{(\beta-1/2)/\gamma}$.
If the local system is described
by mean-field theory, then $\beta-1/2=0$.  In other cases,
$\beta<1/2$, and the mean-field amplitude $\tilde A$ diverges for small $p$.
From Eq.~(\ref{mfa}),
the specific heat jump at $T=\tilde T_c$ is equal to
$(\gamma A_{\chi}^{+} |\tilde T_c-T_c|^{-\gamma-1})^2
/(2B_{\chi} |\tilde T_c-T_c|^{-\gamma-2\delta \beta})\propto
|\tilde T_c-T_c|^{2\delta\beta-\gamma-2}=|\tilde T_c-T_c|^{-\alpha}$.
Thus, the jump in specific heat at the mean-field transition is of order
the specific heat of the local system at temperature $\tilde T_c$.

We now consider the width of the
mean-field critical region.
If we extrapolate Eq.~(\ref{mtct}) to $T=T_c$, we find that the power
of $p$ in the result is consistent with the power of $p$ in Eq.~(\ref{mtc}).
For $T<T_c$, 
the average
magnetization of the local system in the absence of a field behaves as
$|T_c-T|^{\beta}$, and the susceptibility is given by
$\chi=A_{\chi}^- |T_c-T|^{-\gamma}$.
In the long-range system,
this magnetization produces a field $h\propto p |T_c-T|^{\beta}$, which
in turns feeds back and increases the magnetization an amount of
order $p |T_c-T|^{\beta} |T_c-T|^{-\gamma}$.  For
$|T_c-T|^{\gamma}>>p$, this effect is negligible compared to
the averaged field itself, $|T_c-T|^{\beta}$.  Thus, at such temperatures
the long-range interactions have negligible effect on the magnetization 
and so the mean-field critical behavior only extends to $|T_c-T|\propto
p^{1/\gamma}\propto |\tilde T_c-T_c|$.  Therefore, for small $p$, the
width of the mean-field critical region is small.  The scaling arguments
above all rely on this width becoming narrower
than the width of the anomalous
critical region in the local system, in which case both
mean-field and anomalous scaling will be seen in the same
system.

{\it Effect of Randomness--}
We now consider the difference between the long-range model, which
lacks randomness and has links of {\it strength} $p/V$, and a version of the
small-world model in which
the local network is modified by adding strong links connecting sites
with a {\it probability} $p/V$.  The strategy is to consider the system
without randomness, and to consider the effects of randomness as
a perturbation.  We verify that it is self-consistent to ignore randomness,
at least for small $p$.

At $\tilde T_c$, the correlation
length of the local system is $\xi\propto (\tilde T_c-T)^{-\nu}\propto
p^{-\nu/\gamma}$.  Thus, within a correlation volume, there are
$p^{-\nu d/\gamma}$ sites.  The correlation volume can
be characterized by the average field acting on it, and the average
temperature of the correlation volume.

First consider the average field.
If sites in the correlation volume are chosen
instead with probability $p$ to have long-range links, then an average
of $p^{1-\nu d/\gamma}$ sites are chosen.  Since $1-\nu d/\gamma<0$,
this number diverges
as $p\rightarrow 0$.  Then, there are a large number of such sites within each
correlation volume and so the
fluctuation in the number of such sites within each correlation volume is
negligible.  Thus, the fluctuation in the average field is negligible.

Now, consider the average temperature.  A site that is coupled via a
long-range link has reduced correlations with its neighbors compared to
one which does not have such a link.  The site it is coupled to via the 
long-range link produces an average field proportional to $m$,
and also produces fluctuations about this field.
Thus, effectively the temperature of a site with a long-range link
is raised, as the fluctuations in the field reduce its correlations
with its neighbors.  Consider the number of sites with long-range links.
The root mean square
fluctuation in the number of such sites scales as $(p\xi^d)^{1/2}$,
and thus the fluctuation in the temperature averaged over a correlation volume 
scales as $(p/\xi^d)^{1/2}\propto
p^{1/2+\nu d/(2\gamma)}\propto |\tilde T_c-T_c|^{\gamma/2-\nu d/2}$.
Compare this to the difference in temperatures,
$\tilde T_c-T_c$.  As long as
\be
\label{mhc}
\gamma/2+\nu d/2>1,
\ee
the fluctuation in temperature is also negligible.  Eq.~(\ref{mhc})
resembles the Harris criterion\cite{hc} for the
relevance of disorder, 
with an additional term $\gamma/2$ on the left-hand
side.  This additional term guarantees that the inequality holds for
all models that we know.
Thus, fluctuations in field and temperature are both negligible and
the scaling of both $\tilde T_c-T$ and $\tilde A$ with $p$ should be
the same in the small-world and long-range models.

{\it Variational Approach---}
In addition to the scaling arguments above, we now show that
the long-range model without randomness provides an
upper
bound to the free energy of a small world network model, using
an argument inspired by the Migdal-Kadanoff bond-moving
procedure\cite{mk}.  Eq.~(\ref{sl})
defines the statistical weight $S$
without randomness.  If instead of connecting every
pair of sites with strength $p/V$, we add connections between
pairs of sites with probability
$p/V$ and unit strength, we obtain a new statistical weight
\be
\tilde S[\phi]=
S_{\rm local}[\phi]-\frac{1}{2}\sum\limits_{x_1,x_2} \phi(x_1) \cdot
\phi(x_2),
\ee
where the sum extends over sites $x_1,x_2$ which are connected by
a long range link. 
Then, ${\langle S-\tilde S \rangle}=
\sum_{x_1,x_2} w(x_1,x_2) \langle \phi(x_1)\phi(x_2)\rangle$,
where the brackets
$\langle\rangle$ denote an expectation value computed with
statistical weight $e^{-S}$, and where $w(x_1,x_2)=-p/(2V)$ if
$x_1,x_2$ are not connected in the small world
network and $w(x_1,x_2)=1-p/(2V)$ if $x_1,x_2$
are connected.  
The quantity ${\langle S-\tilde S \rangle}$ is a random function of
disorder.  However, for a large system, this quantity is
self-averaging.  If pairs are connected with probability $p/V$,
the average over disorder, $\overline{\langle S-\tilde S \rangle}$, is
equal to zero.  Thus, for
typical networks, $\langle S-\tilde S\rangle=0$, up to small fluctuations,
and thus by a convex inequality, $\tilde Z_{\rm typ}\geq Z$, where
$\tilde Z_{\rm typ}$ is the partition function for a typical small world
network, with statistical weight $e^{-\tilde S}$.

This shows that the free energy of such a small world model is less than
or equal to the free energy of the long-range model.
Intuitively, we expect that the transition temperature in the
long-range model will be higher than that in the small-world model: if
we ignore the local couplings, and consider only the long-range links
then this statement is definitely true.
Above we have argued that $\tilde T_c-T$ should
scale as the same power of $p$ in both the small-world and long-range models.
Even if that argument were to break down, we should expect that
$\tilde T_c-T_c$
scales as at least as large a power of $p$ in the small-world model as
it does in the long-range model.

{\it Comparison to Numerics---}
An important work was a numerical calculation of some of these
quantities, looking for the shift in the transition temperature\cite{eq4}.
In that paper, a different scaling argument was made for the shift,
$\tilde T_c-T_c\sim p^{1/(\nu d)}$.  
This is the temperature
at which a correlation volume includes roughly one long-range link.
However, we have argued that the shift in transition
temperature actually scales as $p^{1/\gamma}$, which is less than
$p^{1/(\nu d)}$ as $p\rightarrow 0$.  The difference
arises since one long-range link is not sufficient to
magnetize an entire correlation volume; several such links are required.

The numerical results in two dimensions are consistent with a
shift in transition temperature scaling as $p^{1/\gamma}=p^{0.57...}$.  The
numerical results in three dimensions indicate
a shift scaling as $p^{0.96}$, while taking $\gamma=1.2396$ from
$\epsilon$-expansion\cite{zj} gives $1/\gamma \approx 0.81$.  This indicates
some discrepancy with the numerical results.  However, in the numerical
study\cite{eq4}, it was argued that their results do not yet involve
sufficiently large lattices to obtain accurate scaling; certainly,
$p^{0.81}$ is closer to the observed scaling than $p^{1/(\nu d)}\approx
p^{0.53}$ is.

{\it Non-Equilibrium Dynamics---}
We now consider the generalization to a non-equilibrium process, the
contact process\cite{teh}, in which each site
is marked either infected or susceptible.  An infected site becomes
susceptible at unit rate, while an infected site can turn a neighboring
susceptible site infected at a rate $\lambda/q$, with $q$ the lattice
coordination number.  The state with all sites susceptible is absorbing.
However, above a critical $\lambda_c$, if a single infected site is
placed in an infinite lattice of susceptible sites, there is a non-zero
probability of the epidemic persisting for all time.
We modify the model as follows: each susceptible site can be infected
by any other infected site, not necessarily a neighbor, at a rate
equal to $p/V$.  Although we will not decouple this interaction,
the general development will be very similar to the equilibrium
case.

We start by recalling some exponents in the local case.
For $\lambda>\lambda_c$,
there is an average density, $\rho\propto |\lambda-\lambda_c|^{\beta}$.
In the presence of a source, where susceptible sites become infected
at a rate $h$, the density $\rho(\lambda=\lambda_c,h)=A_{\rho}
h^{1/\delta_h}$.   

Consider also the infection spreading from a single source.  At
$\lambda=\lambda_c,$ the
survival probability of the infection after time $t$, $P(t)$ obeys
$P(t)=A_p t^{-\delta}$,
The number of infected sites is a random variable, $n(t)$; the
average number of such sites obeys $\overline n(t)=A_n t^{\eta}$.  
The radius of the infection
scales as $t^{z/2}$.
For $\lambda<\lambda_c$, the infection dies out exponentially,
with an asymptotic survival probability $P(t)\propto e^{-t/\tau}$,
with $\tau\propto |\lambda_c-\lambda|^{-\nu_{\|}}$.  This gives
rise to a divergent
susceptibility: in the presence of a source $h$, the susceptibility,
$\chi\equiv\partial_h \rho$, obeys
$\chi(\lambda,h=0)=A_{\chi} |\lambda_c-\lambda|^{-\gamma}$.

Now, consider the dynamics in the long-range model, with $\lambda=\lambda_c$,
with a single source for an infection.
This source grows as described, with the
given $P(t),\overline n(t)$.  However, the local outbreak 
starting from that source 
can produce other local outbreaks elsewhere, via the long-range links, at
a
rate equal to $p$ times the number of infected
sites.   For $p$ small, the number of infected sites $n(t)$
will be large before such
an event, and thus the fluctuations in the $n(t)$ are described
by a random process with a universal distribution.
In the large $V$ limit, at fixed $t$, each new local outbreak produced
via a long-range link is well separated in space from the other local outbreaks.
Thus, we can describe the dynamics of the spread from a single source
simply: there is initially one local outbreak, created at time
$0$, which survives
at time $t$ with probability $P(t)$, and which produces
additional local outbreaks at a rate equal to $pn(t)$.  
Each local outbreak, created at time $t'$, 
evolves independently, surviving with probability $P(t-t')$,
and producing additional local outbreaks with rate $pn(t-t')$.  This fully
describes the dynamics via a branching process.  For $\lambda\neq \lambda_c$,
this description of the dynamics remains valid with a changed 
$P(t)$ and distribution of $n(t)$.

At short times, the average number of infected sites in this dynamics is equal
to $\overline n(t)$.  At long times, the average number of infected sites
grows exponentially.  To describe this exponential growth, realize that
at long times the number of local outbreaks becomes large.  If $s(t')$ describes
the number of local outbreaks started at time $t'$, then the average number
of particles at time $t$ is equal to $\int_0^{t} {\rm d}t' \, s(t') n(t-t')$,
and thus on average $s(t)=p\int_0^{t} {\rm d}t' \, s(t') n(t-t')$.
Inserting an ansatz $s(t)=e^{\alpha t}$, we find that
\be
\alpha=
[A_{\eta} p \Gamma(1+\eta)]^{1/(1+\eta)}.
\ee

Each local outbreak takes a volume of order $t^{dz/2}$.
Eventually, at sufficiently large time, such that 
$e^{\alpha t}\sim (V/t^{dz/2})$,
the individual local outbreaks start to merge, and the dynamics of different
local outbreaks become coupled.  This time $t$ is of order ${\rm ln}(V)$.

Beyond this time $t$, one approaches a stationary state with density $\rho$.
At $\lambda=\lambda_c$,  the dynamics is equivalent to the local system
with a source of particles $h=p \rho$.  Thus, we find
that the density obeys $h=p A_{\rho} h^{1/\delta_h}$, or
$h=(p A_{\rho})^{\delta_h/(\delta_h-1)}$ and thus
\be
\label{rd}
\rho=A_{\rho} (p A_{\rho})^{1/(\delta_h-1)}.
\ee

Eq.~(\ref{rd}) should be compared to Eq.~(\ref{mtc}).  It implies that
the transition to a spreading epidemic happens at 
$\lambda=\tilde \lambda_c<\lambda_c$.  For $\lambda\approx\tilde\lambda_c$,
following the same steps
as in the equilibrium case
leads to the same result as Eq.~(\ref{mtct}), except that
the role of temperature $T$ is replaced by $\lambda$, and the role of
magnetization $m$ is replaced by density $\rho$.  The transition is
again mean-field.  Thus, the stationary results in this non-equilibrium model
are described by the same scaling theory as in the equilibrium models, while
the spread of infection
starting from a single source
is described by an interesting branching dynamics.

{\it Discussion---}  We have developed a general scaling theory
for describing equilibrium and non-equilibrium system with both
short and long-range interactions.  We find that the long-range
interactions lead to mean-field behavior, but with a scaling region
whose width vanishes as $p\rightarrow 0$.  We have also
developed a branching process description of the spread of infection 
from a single source in the contact process with long-range interactions.

{\it Acknowledgments---}
I thank Z. Toroczkai for discussion and P. Fenimore for a careful reading.
This work was supported by DOE W-7405-ENG-36.

\end{document}